\documentclass[preprint]{aastex}

\shortauthors{Manset and Bastien}
\shorttitle{Polarization of Mie scattering circumbinary envelopes}

\begin{document}

\title{Polarimetric variations of binary stars. II. Numerical simulations
for circular and eccentric binaries in Mie scattering envelopes}
\author{N. Manset\altaffilmark{1} and P. Bastien}
\affil{Universit\'e de Montr\'eal, D\'epartement de Physique, 
and Observatoire du Mont M\'egantic, C.P. 6128, Succ. Centre-ville, 
Montr\'eal, QC, H3C 3J7, Canada}
\authoremail{manset@cfht.hawaii.edu, bastien@astro.umontreal.ca}

\altaffiltext{1}{Now at: Canada-France-Hawaii Telescope Corporation,
P.O. Box 1597, Kamuela, HI 96743, USA} 

\begin{abstract}
Following a previous paper on Thomson scattering, we present numerical
simulations of the periodic polarimetric variations produced by a binary
star placed at the center of an empty spherical cavity inside a
circumbinary ellipsoidal and optically thin envelope made of dust
grains. Mie single-scattering (on spherical dust grains) is considered
along with pre- and post-scattering extinction factors which produce a
time-varying optical depth and affect the morphology of the periodic
variations. The orbits are circular or eccentric. The mass ratio (and
luminosity ratio) is equal to 1.0. We are interested in the effects that
various parameters (grain characteristics, geometry of the envelope,
orbital eccentricity, etc.) will have on the average polarization, the
amplitude of the polarimetric variations, and the morphology of the
variability.

We show that the absolute amplitudes of the variations are smaller for
Mie scattering than for Thomson scattering, which makes harder the
detection of polarimetric variations for binary stars surrounded by dust
grains.

The average polarization produced depends on the grains' composition and
size, and on the wavelength of observation. Among the four grain types
that we have studied (astronomical silicates, graphite, amorphous
carbon, and dirty ice), the highest polarizations are produced by grains
with sizes in the range $a \sim 0.1\,$--$\,0.2 \, \micron$ ($x = 2 \pi a
/ \lambda \sim 1.0\,$--$\,2.0$ for $\lambda = 7000\,$\AA). Composition
and size also determine if the polarization will be positive or
negative.

In general, the variations are double-periodic (seen twice per
orbit). In some cases, because spherical dust grains have an asymmetric
scattering function, the polarimetric curves produced show
single-periodic variations (seen once per orbit) in addition to the
double-periodic ones. A mixture of grains of different sizes does not
affect those conclusions.

Circumstellar disks produce polarimetric variations of greater amplitude
(up to $\sim 0.3$\% in our simulations) than circumbinary envelopes
(usually $\lesssim 0.1$\%). Other geometries (circumbinary flared disks
or prolate envelopes, and non-coplanar envelopes) do not present
particularly interesting polarimetric characteristics.

Another goal of these simulations is to see if the 1978 BME (Brown,
McLean, \& Emslie) formalism, which uses a Fourier analysis of the
polarimetric variations to find the orbital inclination for
Thomson-scattering envelopes, can still be used for Mie scattering. We
find that this is the case, if the amplitude of the variations is
sufficient and the true inclinations is $i_{\rm true} \gtrsim
45\arcdeg$. For eccentric orbits, the first-order coefficients of the
Fourier fit, instead of second-order ones, can be used to find almost
all inclinations.

\end{abstract}


\section{Introduction}

In the first paper (Manset \& Bastien 2000, hereafter referred to as
Paper~I), we presented numerical simulations of the periodic
polarimetric variations produced by a binary star surrounded by a
circumbinary envelope composed of electrons. Thomson scattering was
considered in an optically thin situation, along with pre- and
post-scattering extinction factors which produce a time-varying (small)
optical depth\footnote{Note: The optical depth has to be small to ensure
that the assumption of single scattering in the envelope is not
violated.} and affect the morphology of the periodic variations by
adding an additional harmonic to the variations. We studied the effects
that orbital inclination, optical depth, geometry of the envelope and
cavity, size and eccentricity of the orbits, and non-equal mass stars
had on the average polarization level and the amplitude of the
polarimetric variations.

We found that high polarization levels will result from a high
inclination, a high optical depth, a flat envelope, or a big central
cavity. Polarimetric variations are more apparent for a low inclination,
a high optical depth, a flat envelope, a small cavity, or an orbit which
brings the stars close to the inner edge of the cavity.

Using the formalism developed by Brown, McLean, \& Emslie (1978,
hereafter BME) the observations, or in this case the results of the
numerical simulations, are fitted with a Fourier series with
coefficients of orders zero, one, and two (see Equations~3 and 4
below). The first order coefficients from the fit (terms in $1 \lambda$)
correspond to single-periodic variations (variations seen once per
orbit), and those of second order (terms in $2 \lambda$) to
double-periodic variations (seen twice per orbit). Those coefficients
are used to find the orbital inclination (see Equations~5 and 6 below).

It was then shown that the BME formalism can be used to find the orbital
inclination if it is $\gtrsim 45\arcdeg$, even though this formalism
does not include variable absorption effects as in our simulations. The
geometry (flatness of the envelope, size of the central cavity) and size
of the orbit have no significant influence on the inclination found by
the BME formalism.

An extension of the BME formalism for eccentric orbits has been
performed by Brown, Aspin, Simmons, \& McLean (1982); however, their
case considers a localized scattering region in eccentric orbit about a
star, which is not a suitable geometry for the cases we want to study
here. Therefore, we performed our own study of eccentric orbits.

For eccentric orbits, single-periodic variations appear for
eccentricities as low as 0.10. As the eccentricity increases, these
single-periodic variations dominate over the double-periodic ones. For
low eccentricities, $e \lesssim 0.3$, the inclinations can be found with
the first or second-order Fourier coefficients, if $i>20\arcdeg$ and
$i>45\arcdeg$ respectively. For the high eccentricities, $0.3 < e <
0.6$, only the first-order coefficients should be used, if
$i>10\arcdeg$.

Orbital eccentricity is not the only factor that can introduce
single-periodic variations: variable absorption effects, non equal mass
stars, and asymmetric envelopes may also produce those.

In this second paper, we investigate the effects of scattering by
spherical dust grains (Mie scattering). In addition to studying the
influence of the optical depth and true orbital inclination already
considered in Paper~I, we also consider the effects of the composition and size
of the dust grains on the level of polarization and the amplitude of the
polarimetric variations. Our goal is to study the periodic polarimetric
variations of binary young stars surrounded by envelopes in which dust
grains are responsible for the scattering and polarization, and compare
polarimetric observations with our numerical simulations.

As we have shown in Paper~I that the BME formalism is still valid beyond
its original limits (circular orbits, no extinction effects), now we
also want to see if the BME formalism can be extended to the case of Mie
scattering to find the orbital inclination, even though electrons and
spherical dust grains have different scattering properties, and the BME
formalism has been developed for Thomson scattering.

The simulations presented here are akin to the simulations where
single-periodic polarimetric variations, usually double-peaked, are
produced by a spotted star surrounded by a dusty circumstellar disk (see
for e.g. Stassun \& Wood 1999). Whereas in the latter case the
polarimetric variations are produced by a pair of intrinsically
polarized hot spots illuminating (and scattering off) the disk, here the
variability is produced by the binarity of the source of light. The
polarization produced by a single star surrounded by a circumstellar
disk has been studied with Monte Carlo simulations, by, for e.g., Wood
et al. (1996).

\section{The scattering model}
The scattering model used to compute the polarization and its position
angle produced by two stars in orbit at the center of an ellipsoidal
envelope was presented in Paper~I. In addition to the parameters
presented there (geometry of the envelope, orbit characteristics,
optical depth, grid size), here we have to choose a grain composition
(which will give a specific complex refractive index) and the size of
the grains. Table~\ref{Tab-grains} presents the characteristics of the
grains studied in this paper. A uniform size distribution was used for
most of our simulations, but we also studied a distribution of grain
sizes.

The calculations for Mie scattering use the following formulas:

\begin{eqnarray} 
Q &=& \frac{\displaystyle \tau_0 e^{\tau_0} 
\sum\, (\cos 2\beta)(i_2-i_1) \, e^{-\tau_1-\tau_2} \, dv/r^2 * iw}
{\displaystyle \tau_0 e^{\tau_0} \sum\, (i_1+i_2) \,e^{-\tau_1-\tau_2} 
\, dv/r^2 +
2\pi \left( \frac{2\pi a}{\lambda} \right)^2 Q_{\rm ext}
\, \frac{x_{{\rm n}0}}{n_{\rm zmax}}}, \end{eqnarray}

\begin{eqnarray} 
U &=& \frac{\displaystyle \tau_0 e^{\tau_0} 
\sum\, (\sin 2\beta)(i_2-i_1) \, e^{-\tau_1-\tau_2} \, dv/r^2 * iw}
{\displaystyle \tau_0 e^{\tau_0} \sum\, (i_1+i_2)e^{-\tau_1-\tau_2} dv/r^2 +
2\pi \left( \frac{2\pi a}{\lambda} \right)^2 Q_{\rm ext}
\, \frac{x_{{\rm n}0}}{n_{\rm zmax}}}, \end{eqnarray}
where:\\
$\tau_0$: optical depth in the equatorial plane of the envelope;\\
$\tau_1$: optical depth before scattering (between the light source and the
scatterer);\\ 
$\tau_2$: optical depth after scattering (between the scatterer and the
observer);\\ 
$\beta$: angle between the East--West direction and the scattering
plane (which includes the light source, the scatterer and the observer);\\
$\psi$: scattering angle, between the light source, the scatterer and
the observer;\\
$r$: distance between the light source and the scatterer;\\
$dv$: volume element (computed in relation to the grid size);\\
$iw$: weight which equals 1 if there is a scatterer, 0 otherwise;\\
$\frac{x_{{\rm n}0}}{n_{\rm zmax}}$: optical depth per grid step;\\
$a$: radius of the spherical dust grain;\\
$\lambda$: wavelength of observation;\\
$Q_{\rm ext}$: extinction cross section;\\
$i_1$ and $i_2$: van de Hulst intensities.\\

Variable absorption effects are considered by using $\tau_1$ and
$\tau_2$. Depending on where the binary star is located inside the
envelope, the optical depths between the source and the scatterer
($\tau_1$), and between the scatterer and the observer ($\tau_2$) will
vary, and with them, the polarization. The van de Hulst intensities and
$Q_{\rm ext}$ are calculated with the usual formulas that can be found
in van de Hulst (1981, chap. 9) and make use of the complex refractive
index of the grains.

The polarization $P$ and its position angle $\theta$ are calculated from
the Stokes parameters $Q$ and $U$. When the polarization is produced by
scattering, the sign used for $P$ is related to its orientation with
respect to the scattering plane (the plane containing the source of
light, the scatterer, and the observer). A positive polarization is
perpendicular to the scattering plane, whereas a negative one is
parallel to it. Thomson scattering always produces positive
polarization, but both positive and negative polarization are produced
by Mie scattering.

As in Paper~I, a ``canonical simulation'' is a simulation which has the
following parameters: an envelope with axis ratios 1.0, 1.0 and 0.25
(also referred to as a 25\% flat envelope), a spherical cavity with
radius 0.20 (also referred to as a 20\% cavity), an optical depth of
0.1, and a grid size of 65. In addition, calculations are made for a
wavelength of $7000\,$\AA. Figure~\ref{Fig-CB_2_05_05} presents a
schematic view of this canonical geometry.

For further details on the calculations, see Paper~I.

\section{The BME formalism}
The simulated polarimetric curves produced by the scattering model
presented above are analyzed with the BME formalism, as in Paper~I. For
reference, we give the main formulas that are used. Observations are
represented as first and second harmonics of $\lambda=2\pi\phi$, where
$\phi$ is the orbital phase ($0.0 < \phi < 1.0$):\\
\begin{eqnarray}
Q &=& q_0 + q_1 \cos \lambda + q_2 \sin \lambda + q_3 \cos 2\lambda +
q_4 \sin 2\lambda, \\
U &=& u_0 + u_1 \cos \lambda + u_2 \sin \lambda + u_3 \cos 2\lambda +
u_4 \sin 2\lambda. 
\end{eqnarray}

The inclination can be found with the first (Equation \ref{EQ-iO1-p2})
or second (Equation \ref{EQ-iO2-p2}) order Fourier coefficients,
although it is expected for circular orbits that second order variations
will dominate:\\

\begin{eqnarray}
\left[ \frac{1-\cos i}{1+\cos i} \right]^2 &=& \frac{(u_1+q_2)^2 +
(u_2-q_1)^2}{(u_2+q_1)^2 + (u_1-q_2)^2} \label{EQ-iO1-p2},\\
\left[ \frac{1-\cos i}{1+\cos i} \right]^4 &=& \frac{(u_3+q_4)^2 +
(u_4-q_3)^2}{(u_4+q_3)^2 + (u_3-q_4)^2}. \label{EQ-iO2-p2}
\end{eqnarray}

In an alternative representation, the eccentricity of the ellipse in the
$QU$ plane is related to the inclination. For the $2\lambda$ ellipse:\\
\begin{equation} e = \frac{\sin^2i}{2-\sin^2i}. \end{equation}
For the $1\lambda$ ellipse:\\
\begin{equation} e = \sin i. \end{equation}

An important difference between this BME formalism and our numerical
simulations is the inclusion of variable absorption effects in our work.

%
%
%
%

\section{Mie Scattering, circular orbits: average polarization and 
amplitude of the polarimetric variations} 

\subsection{\label{grid-size-p2}Choice of grid size}
As in Paper~I, a preliminary step to the numerical calculations was to
determine a suitable grid size, one with which the BME formalism would
find an orbital inclination close enough to the real inclination (say,
within $\approx 2 \arcdeg$). Using astronomical silicates grains with
radii of $0.1 \, \micron$, we calculated models with grid sizes (radius)
25 to 95, by steps of 10. The results of the simulations, in which no
noise was introduced, were used as input data for the BME equations.
The orbital inclinations found in this manner are almost identical
(within a few tenths of a degree) to those we found for Thomson
scattering in Paper~I. As in Paper~I, we chose a grid size of 65, for
which the inclination is within $1 \arcdeg$ of the true value, and the
calculations can be performed in reasonable times.

\subsection{Effects of the composition and size of the spherical dust
grains} We made calculations for the canonical simulation (25\% flat
envelope, 20\% central cavity, $\tau=0.1$, grid radius=65), the four
grain compositions listed in Table~\ref{Tab-grains}, and five grain radii
(0.02, 0.05, 0.10, 0.2, and $0.50 \, \micron$), at a wavelength of
$7000\,$\AA. Since the polarization produced by Mie scattering depends
on the size of the grain $a$ and the wavelength of observation
$\lambda$, it is common to introduce the parameter $ x = 2\pi a /
\lambda$; the values of $x$ for the previously mentioned simulations are
0.18, 0.45, 0.90, 1.8, and 4.5. For each simulation, we used only one
grain size and one grain composition.

If we compare the polarizing properties of dust grains and electrons, we
note that dust grains (irrespective of composition and size) are less efficient
polarizers than electrons. Grains also produce smaller absolute
polarimetric variations per particle than electrons, although both
electrons and grains produce about the same $\Delta P / P$ ratios; see
Table~\ref{Tab-grains-e}. That important observation shows that for
binary stars embedded in dust envelopes, periodic polarimetric
variations should be harder to observe and detect.

For grain sizes of 0.02, 0.05 or $0.10 \, \micron$ ($x=0.18, 0.45,
0.90$), the highest polarization and polarimetric variations are found
for astronomical silicates, followed by graphite and amorphous carbon,
and finally dirty ice; grains made of dirty ice are 5--10 times less
polarizing and produce variations about 5--10 times smaller than grains
made of astronomical silicates. For grains with $a=0.20 \, \micron$
($x=1.8$), the dirty ice produces the highest polarization and
variations. For $a=0.5 \, \micron$ ($x=4.5$), astronomical silicates,
graphite, and amorphous carbon have similar polarimetric properties, and
dirty ices are less efficient polarizers. The highest polarizations are
produced by grains with sizes in the range $\sim 0.1$--$0.2 \, \micron$
($x \sim 1.0$--2.0).

For astronomical silicates, we also did simulations for a wider range of
grain sizes, going from $0.01 \, \micron$ to $0.55 \, \micron$. The
results are presented in Figure~\ref{Fig-P_vs_x}, where it is seen that
the maximum polarization is reached for $x \approx 1.0$.

These results follow what is presented in Simmons (1982). See for
example his figure~4 (polarization as a function of $x$ for different
grain compositions), where it can be seen that for silicate grains and ice,
polarization depends on $x$ with a very different behavior for these 2
grain compositions; a grain that is the most polarizing at one value of $x$
will not necessarily be the most polarizing grain at other values of
$x$.

Figure~6 in Simmons (1982) shows for silicate grains in particular how a
different absorption coefficient changes the behavior of the
polarization as a function of $x$. Finally, figure~5a shows that $x_{\rm
max}$, the value of $x$ for which the polarization is maximum, depends
on both the real and imaginary parts of the refractive index (and thus
on the grain composition).

We also made some more calculations for astronomical silicates, adding
grain sizes of $1.0 \, \micron$ and $2.0 \, \micron$. Big grains ($a \ge
0.5 \micron$) are less efficient polarizers, as was already known (see
for example Daniel 1978, Simmons 1983).

As shown by our simulations and by Simmons (1982), the average
polarization produced by dust grains thus depends on the grain
composition and size.

\subsection{Polarization reversal}

Polarization reversal (when the polarization goes from positive, or
perpendicular to the scattering plane, to negative, or parallel to the
scattering plane) seen in our simulations also occurs for the $x$ values
presented in Simmons (1982), figure~5b. For astronomical silicates,
polarization reversal occurs at a value somewhere between
$a=0.19\,\micron$ and $a=0.25\,\micron$, or $x\approx1.8$ (see
Figure~\ref{Fig-P_vs_x}), in agreement with figure~5b of Simmons (1982).

For dirty ice, we do not see any polarization reversal, which follows
the figure~5b of Simmons (1982), except for our biggest grains
($a=0.5\micron$, $x\approx4.5$) at low inclinations ($10\arcdeg \leq i
\leq 30\arcdeg$). For the grains with large imaginary refractive index
values (graphite and amorphous carbon), we see from table~1 of Daniels
(1980) that polarization reversal is not expected at all, which is what
is seen here, except in 2 cases for intermediate inclinations.

\subsection{Distribution of grain sizes}

We also studied a distribution of grain sizes, using the MRN size
distribution $N(a) \sim a^{-3.5}$ (Mathis, Rumpel, \& Nordsiek 1977) for
astronomical silicates with sizes between $0.01\, \micron$ and $0.50 \,
\micron$. The results are not very different from the case where only
one grain size was considered. Since all polarimetric curves have the
same morphology for a given set of parameters except grain size, adding
many of those curves only changes the polarization level, not the
morphology.

\subsection{\label{i_on_P-p2}Effect of true orbital inclination}

As the inclination decreases from $90^\circ$ (edge-on) to $0^\circ$
(pole-on), the polarization decreases, in general following a
theoretical law (developed for a single star at the center of an
axisymmetric envelope but nonetheless interesting for our case) that can
be found for example in Brown \& McLean (1977): the residual
polarization (sum of polarization of the scattered light and the
unpolarized light from the star) scales as $\sin^2(i)$. Deviations from
this theoretical law are found when the optical depth is high
($\tau=0.5$, too high to consider single scattering only) or the grains
are big ($a\ge0.2\, \micron$, which probably introduces higher frequency
variations in the polarization as a function of the scattering angle).
With a $90^\circ$ inclination, the $U$ parameter is null, and only the
$Q$ parameter varies. As the inclination decreases, both the $Q$ and $U$
parameters vary, and more so as the inclination decreases.  In general,
the behavior is the same as for Thomson scattering.

\subsection{Periodic polarimetric variations}
The simulations for a binary star in a circumbinary envelope give
polarimetric variations that are double sine waves (see
Figure~\ref{Fig-binary_AS_t01_a01}), which translate to an ellipse that
is traced out twice per orbit in the $QU$ plane. The average Stokes
parameter $U$ is zero, as is expected from a geometry oriented in the
plane of the sky in a East--West direction, and the position angle is
usually $\approx 0\arcdeg$, perpendicular to the projection of the
envelope's main axis and scattering angle.

In order to decrease the number of parameters influencing the
polarization and isolate causes and their effects, we can take a look at
the contribution of only one star instead of considering the combined
effects of the 2 stars. In Figure~\ref{Fig-primary_AS_t01_a01} we show
the polarimetric variations for the primary only, as if it were in orbit
around an ``invisible'', or much fainter, secondary star, both stars
being in orbit at the center of the same circumbinary shell. Even though
the orbit is circular, single-periodic variations are seen in addition
to the usual double-periodic ones. We have shown in Paper~I
that variable absorption effects introduce such single-periodic
variations; but if we neglect those variable absorption effects and
impose a constant optical depth in our Mie simulations, those
single-periodic variations persist. This is due to the asymmetric phase
function of dust grains. It is also seen that the peaks of polarization
are not located exactly at phases 0.25 and 0.75, as expected for an
asymmetric phase function.

Therefore, in addition to variable absorption effects, non equal mass
stars, orbital eccentricity, and asymmetric envelopes, Mie scattering can
also cause single-periodic variations.

\subsection{\label{orbits-geom-p2}Effect of orbits and various geometries}
Like the Thomson scattering results presented in Paper~I, the amplitude
of the variations increases significantly for larger size orbits. For an
orbit that comes close to the inner edge of the circumbinary disk, the
polarimetric variations are more apparent.

In order to see the effects of geometries different than a circumbinary
ellipsoidal envelope, we made simulations for flared circumbinary disks,
prolate circumbinary envelopes, and circumstellar disks. We also
investigated the effects of a non-coplanar geometry where the plane of
symmetry of the envelope and the orbital plane differ from one another.

First, we tried a geometry similar to a flared disk, with a conic
opening (with half-angle between 10 and $80 \arcdeg$) in an ellipsoidal
envelope. As the angle of the cone increases, so does the polarization
and amplitude of the variations. We also studied prolate geometries
instead of oblate ones, orienting the long axis of the envelope along 3
axes (toward the observer, in a N--S and E--W orientation in the plane
of the sky). Finally, we inclined the envelope by $\pm 30 \arcdeg$ with
respect to the orbital plane. For all those cases, polarization and
amplitude of variations are affected to various degrees, but the
polarimetric variations have the same morphology.

For circumbinary envelopes, the polarimetric curves we have produced so
far seldom have an amplitude greater than 0.10\%, whereas Simmons (1983)
could produce amplitude of many tenths of a percent, or even of a few
percents, with a geometry more favorable for high amplitude polarimetric
variations: a single circumstellar envelope externally illuminated by
only one star.

Indeed, the most interesting simulations were done with a circumstellar
disk around the primary, a ``naked'' secondary which does not have any
circumstellar material in its environment, and no circumbinary disk.
See Figure~\ref{Fig-CS_45_1_5} for an example of such a geometry. As
expected from this more asymmetric geometry, the polarimetric variations
produced have a greater amplitude, ranging from 0.1 to 0.3\%.  The
primary star that sits in the middle of its circumstellar disk does not
produce any variation, as expected, but numerical noise is present at a
level of $\lesssim 0.01$ \% in polarization. Variations for the
secondary star, which revolves around the primary and its disk, are a
sum of double- and single-periodic variations.

\section{Mie Scattering, circular orbits: finding the orbital inclination}

Average polarization, amplitude of the polarimetric variations, and
morphology of the variability are all interesting, but periodic
polarimetric observations can potentially give an additional and very
interesting parameter for a binary star: its orbital inclination, which
is found by using the formalism developed by BME
(Equations~\ref{EQ-iO1-p2} and \ref{EQ-iO2-p2}). We now investigate if
their formalism, which was developed for Thomson scattering, circular
orbits, and constant optical depth, can be extended successfully to our
simulations to find the orbital inclination.

\subsection{Effects of true orbital inclination, composition and size of the 
spherical dust grains on the orbital inclination found by the BME
formalism} 
For an optical depth of 0.1, we have found that the BME formalism will
work (i.e., true inclinations can be found from the BME analysis of the
polarimetric variations if $i_{\rm true} \gtrsim 45\arcdeg$) as long as
there are polarimetric variations of sufficient amplitudes for the
mathematical analysis (by sufficient, we mean for example, 0.002\% for
simulations without stochastic noise). The composition or the size of
the grains is not important as long as some variations are
introduced. Big grains ($a \gtrsim 0.5\, \micron$) do not produce enough
variations for the BME analysis to work properly.

In general, the uncertainty on the inclination found by the BME
formalism (calculated with the method of propagation of errors) is
higher for dust grains than for electrons. This is understandable since,
as stated before, grains produce smaller amplitude polarization
variations, which will increase the uncertainty on the inclination
found.

\subsection{\label{tau-p2}Optical depth effects}
The effects of optical depths were studied for some combinations of
grain composition and size, with optical depths of 0.02, 0.05, 0.2, and
0.5. As expected, the polarization increases with the optical depth, as
does the amplitude of the polarimetric variations. All the simulations
studied here produced polarimetric variations with sufficiently large
amplitudes, so again, the BME formalism works for $i_{\rm true} \gtrsim
45\arcdeg$.

Berger \& M\'enard (1997) have studied multiple Mie scattering in
circumbinary shells with optical depths up to $\tau = 3.0$, and have
found that even with such a high optical depth, the estimation of
inclination from the BME formalism is still very good.

\subsection{\label{orbits-p2}Effect of orbits and various geometries}

Although the inclination found by the BME equations is not significantly
affected by the geometry in the case of the flared disk geometry, it is
more seriously affected for prolate geometries and non-coplanar orbital
and geometric planes. For the prolate envelope inclined at $60 \arcdeg$,
the two cases where the long axis of the prolate envelope is in the
plane of the sky give inclinations close to $60 \arcdeg$, but for the
long axis coming toward the observer, the inclination found is $10
\arcdeg$ too low. For the same inclination, the envelope inclined by
$\pm 30 \arcdeg$ with respect to the orbital plane gives orbital
inclinations of $54 \arcdeg$ or $43 \arcdeg$, significantly different
than the $59 \arcdeg$ found for the coplanar case.

For the circumstellar disks, BME can find the inclination if it is $\geq
30\arcdeg$. The lower threshold, compared to the circumbinary case,
might be due to the amplitude of the variations, which are greater for
the circumstellar envelope geometry than for the circumbinary envelope
geometry.

\section{\label{ecc-p2}Mie scattering, eccentric orbits}

To investigate the possible effects of non-circular orbits on
polarimetric observations, canonical simulations were performed with
orbital eccentricities of 0.1, 0.3, and 0.5. It should be noted that in
these simulations, the orbital semi-major axes have been adjusted
according to the eccentricity (i.e., decreased with increasing $e$) so
the stars came at about the same distance from the inner edge of the
envelope, irrespective of the eccentricity of their orbit (the maximum
distance between the center of the envelope and any of the two stars was
always $\approx0.15$).  This was done so we could study the influence of
the eccentricity without having to deal with the influence of how close
the stars come to the inner edge of the circumbinary envelope.  The
results found for Mie scattering are identical to those we have found
for Thomson scattering (see Paper~I).

The eccentricity has no significant influence on the average level of
polarization, but does change the amplitudes of the polarimetric
variations: the amplitudes decrease as the eccentricity increases.  As
the eccentricity increases, the $2\lambda$ variations that are dominant
for low eccentricities give way to $1\lambda$ variations. Moreover,
higher harmonics appear; a fit with 1$\lambda$ and 2$\lambda$ is not
sufficient to reproduce the polarimetric curves, so 3$\lambda$ and
4$\lambda$ harmonics are needed.

When looking at the inclinations found by the BME formalism with the
second-order coefficients of the fit, we find that the true inclinations
are found for the lower eccentricities ($0.1 < e < 0.3$) and the highest
inclinations ($i_{\rm true} \gtrsim 45\arcdeg$). For the highest
eccentricity studied here, $e=0.5$, the second-order coefficients can
not be used to find the orbital inclination, as the BME formalism is
unable to find the true inclination.

For all eccentricities, even the highest one studied here $e=0.5$, the
first-order coefficients can be used to find almost all
inclinations. There is a numerical problem with the true inclination of
80\arcdeg, for which the BME formalism can not assign the right
inclination. The lowest inclinations ($i_{\rm true} < 20\arcdeg$) are
harder to find.

When the periastron is changed to values other than 0\arcdeg, the
results of the BME analysis are slightly modified. With an eccentricity
of 0.5, the second-order and first-order coefficients now both give
reasonable results for $i_{\rm true} \gtrsim 45\arcdeg$.

\section{Comparison between the numerical simulations and polarimetric
observations of binary young stars} In future papers, we will present
polarimetric observations of binary young stars, which are objects
surrounded by circumstellar material (dust grains). The binaries we have
selected have short periods, so we believe the geometry adopted here
(circumbinary ellipsoidal envelope with a central cavity) is suitable
for these kinds of objects.

Some of the observed pre-main sequence binaries show periodic
polarimetric variations, although they are in general of lower amplitude
and less clear than those of binary hot stars (surrounded by electrons),
such as Wolf--Rayet stars (see, for example, St-Louis et al. 1988;
Drissen et al. 1989b; Robert et al. 1990). This is in agreement with our
previous comparison of the polarimetric properties of electrons and dust
grains.

Non-periodic polarimetric variations that are known to exist even for
single young stars (Bastien 1982; Drissen, Bastien, \& St-Louis 1989a;
M\'enard \& Bastien 1992) produce stochastic noise in the polarimetric
curves of binary young stars, sometimes hiding the low amplitude
periodic variations.

The small amplitude variations ($\lesssim 0.10$\%) often seen in binned
data can be caused by non favorable inclinations or geometry, or not
enough scatterers, but we believe it is in general agreement with one of
our conclusions, that dust grains produce smaller amplitude variations
than electrons.

Some stars do have high amplitude polarimetric variations of up to
0.7\%. Such large amplitudes were not quite produced by our simulations,
although circumstellar disks, a more favorable geometry than
circumbinary disks, do produce more substantial variations, up to $\sim
0.3$\%.  For the observed high amplitude polarimetric variations, a
circumbinary envelope might thus not be an adequate geometry and
circumstellar envelope(s)/disk(s) could be present.

\section{Discussion}
We have presented numerical simulations of the periodic polarimetric
variations produced by a binary star placed at the center of an empty
cavity in a circumbinary ellipsoidal and optically thin envelope. Mie
single-scattering in envelopes made of single-size and
single-composition grains was considered. The orbits were circular or
eccentric.  The mass ratio (and luminosity ratio) was equal to
1.0. These parameters are to represent short-period spectroscopic binary
young stars that are embedded in a circumbinary envelope and have
evacuated the central regions of this envelope due to gravitational or
radiative interactions.

We have shown that the absolute amplitude of the variations is smaller
for Mie scattering than for Thomson scattering, which will render more
difficult the detection of polarimetric variations in binary stars
surrounded by dust grains. In fact, polarimetric observations of binary
young stars that will be presented in future papers show periodic
variations that are sometimes much less obvious than for hot stars, for
example. This may be due in part to the scattering properties of dust
grains, although other factors, such as orbital inclination or
scattering geometry, can also decrease (or increase) the amplitude of
the polarimetric variations.

The average polarization produced depends on the grains' composition
(through the real and imaginary parts of the refractive index) and size,
and on the wavelength of observation; a highly polarizing grain at a
given grain size will not necessarily polarize as efficiently if it is
smaller or bigger. For the four grain compositions that we have studied
(astronomical silicates, graphite, amorphous carbon, and dirty ice), the
highest polarizations are produced by grains with sizes in the range
$\sim 0.1\,$--$\,0.2 \, \micron$ ($x \sim 1.0\,$--$\,2.0$ for $\lambda =
7000\,$~\AA).

Polarization reversal was seen for astronomical silicates (at $a \approx
0.2\, \micron$, or $x \approx 1.8$), but not for dirty ices, graphite,
and amorphous carbon grains, which is in agreement with previous works
(Simmons 1982; Daniels 1980).

The periodic variations produced are in general double-periodic (seen
twice per orbit), although we have shown that variable absorption
effects (pre- and post-scattering factors), non equal mass stars,
orbital eccentricity, asymmetric envelopes, and Mie scattering (on
spherical dust grains, which have an asymmetric scattering function)
introduce single-periodic variations (seen once per orbit).

We have also studied a mixture of grains whose sizes followed a MRN
distribution; this did not affect the morphology of the polarimetric
variations. 

Circumstellar disks are more interesting than circumbinary disks, in the
sense that they produce polarimetric variations that are more readily
detectable ($\sim 0.3$\% versus $\lesssim 0.1$\%). Other geometries
(flared disks, prolate envelopes, non-coplanar disks) did not produce
any difference in the morphology or amplitude of the variations.

We have shown that the BME formalism is still valid beyond its original
limits (circular orbits, no extinction effects, Thomson scattering),
under certain circumstances. In general, the BME formalism will be able
to find the orbital inclination if this inclination is $\gtrsim
45\arcdeg$ and if the amplitude of the polarimetric variations is
sufficient. The composition or the size of the grains is not important as long
as some variations are introduced, but the inclinations found by the BME
formalism seem to be affected by prolate and non-coplanar geometries.

The second-order Fourier coefficients of the fit can be used for the
lower orbital eccentricities ($0.1 < e < 0.3$) and the highest
inclinations ($i_{\rm true} \gtrsim 45\arcdeg$). For all eccentricities,
even the highest one studied here $e=0.5$, the first-order coefficients
can be used to find almost all inclinations.

Comparisons between polarimetric observations of binary stars and these
simulations will help understand these systems (presence of circumbinary
and/or circumstellar disks, geometry of the envelope, orbital
eccentricity, etc.) although the interpretations might not be unique, as
many parameters can have similar effects on the polarimetric variations
(mainly the amplitude and the presence of single-periodic
variations). In that case, other types of observations (imaging and
spectroscopy) would nicely complement polarimetric ones.

If the observations are of high enough quality and show variations of
sufficient amplitude, the BME formalism can be used to determine the
orbital inclination, a fundamental parameter for these stars.

\acknowledgments
N. M. would like to thank the Conseil de Recherche en Sciences
Naturelles et G\'enie of Canada, the Fonds pour la Formation de
Chercheurs et l'Aide \`a la Recherche of the province of Qu\'ebec, the  
Facult\'e des Etudes Sup\'erieures and the D\'epartement de physique
of Universit\'e de Montr\'eal for scholarships, and P. B. for
financial support. N. M. would also like to thank F. M\'enard for
numerous discussions. We would like to thank the Conseil de Recherche en 
Sciences Naturelles et G\'enie of Canada for supporting this research.

\pagebreak

\pagebreak \clearpage


\scalebox{0.75}{\includegraphics{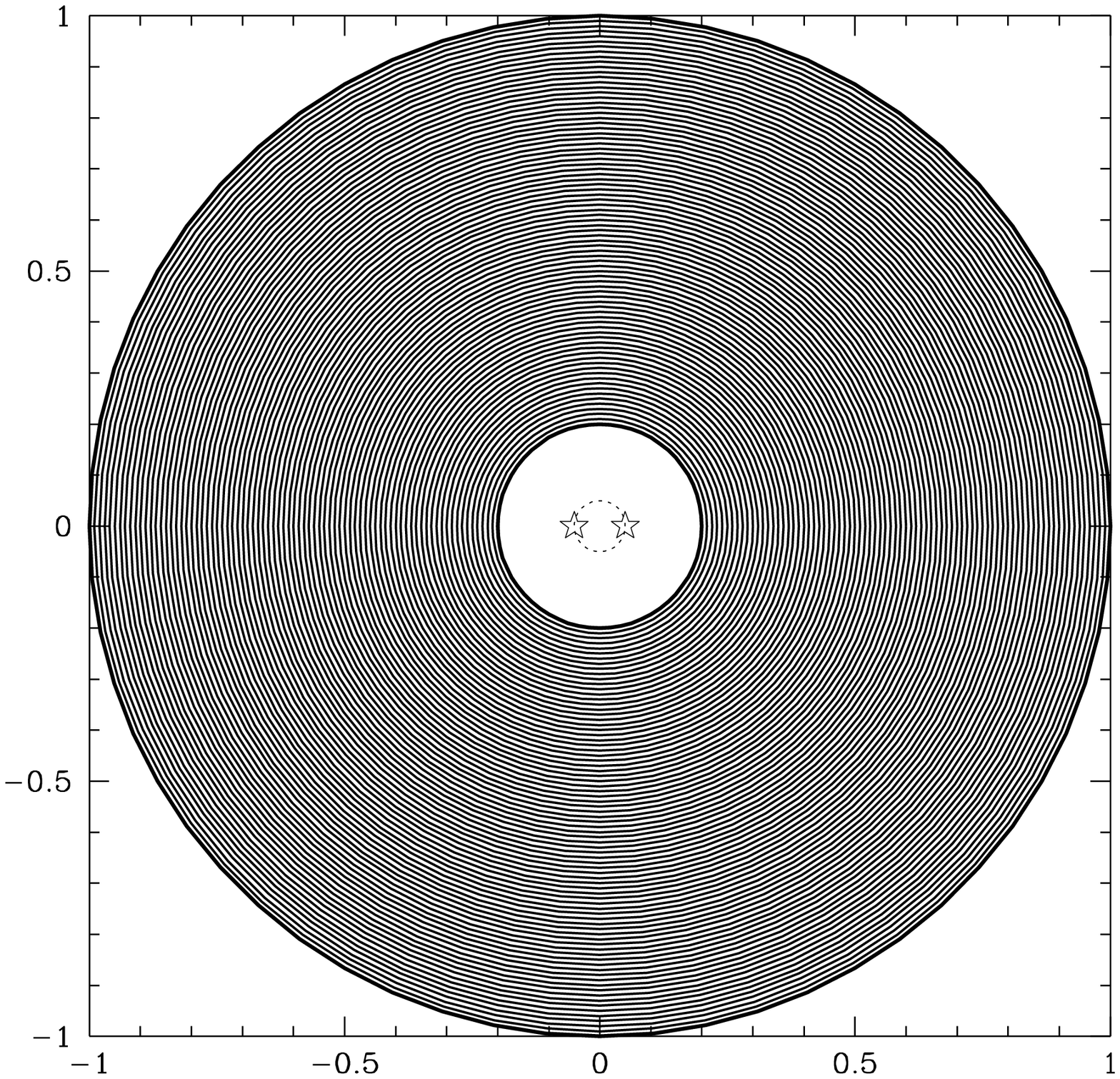}}
\figcaption[CB_2_05_05.ps]{Pole-on view of a circumbinary envelope with
a cavity size 20\% of the envelope size. An equal-mass binary with an
orbital radius 5\% of the envelope size is placed at the center of the
cavity. \label{Fig-CB_2_05_05}}

\newpage
\scalebox{0.75}{\includegraphics{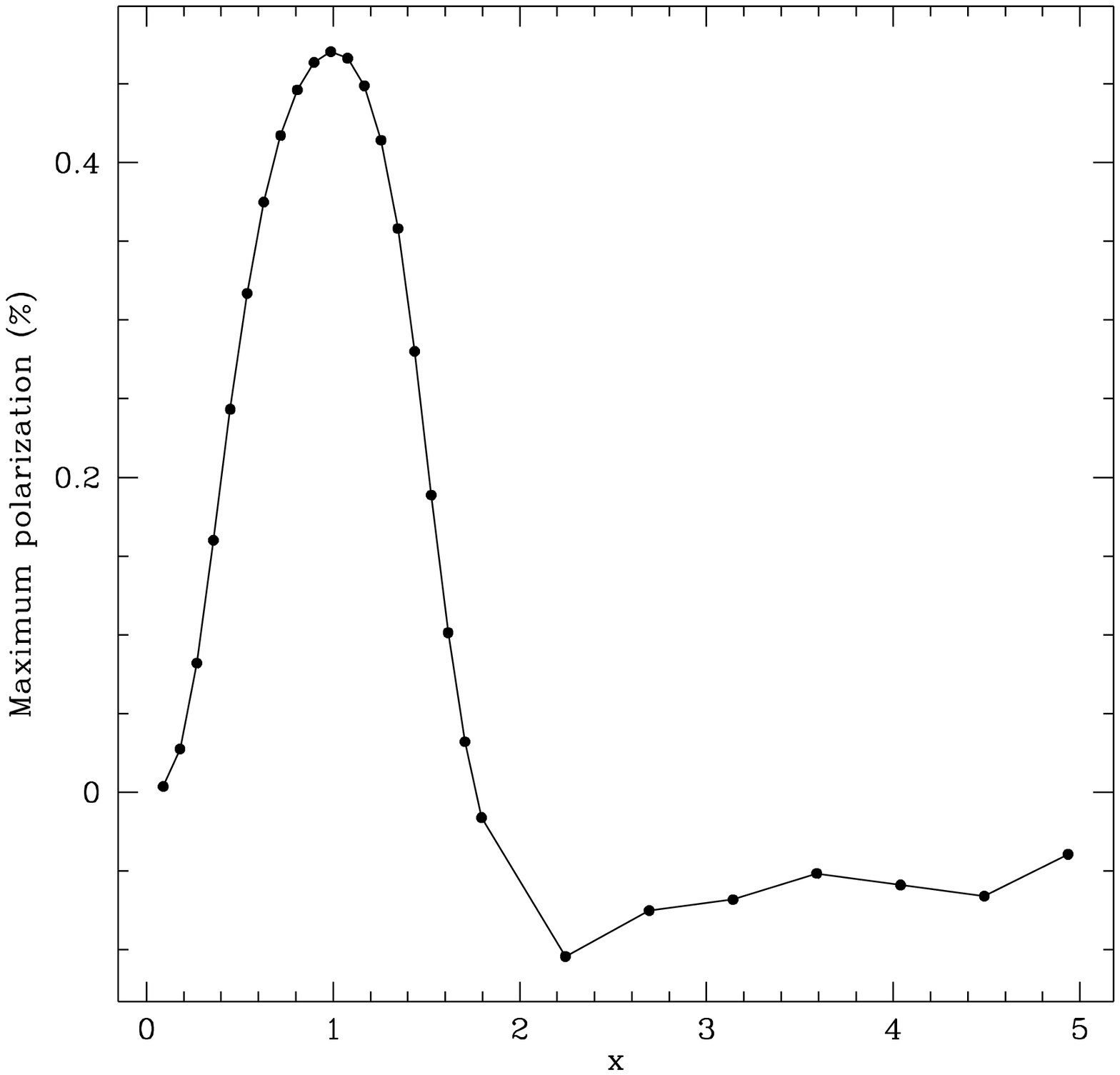}}
\figcaption[P_vs_x.ps]{For astronomical silicate grains, polarization as
a function of $x=2 \pi a / \lambda$. The highest polarization occurs for
$x \approx 1.0$. Polarization reversal occurs at a value somewhere
between $a=0.19\,\micron$ and $a=0.25\,\micron$, or
$x\approx1.8$. \label{Fig-P_vs_x}}

\newpage
\scalebox{0.75}{\includegraphics{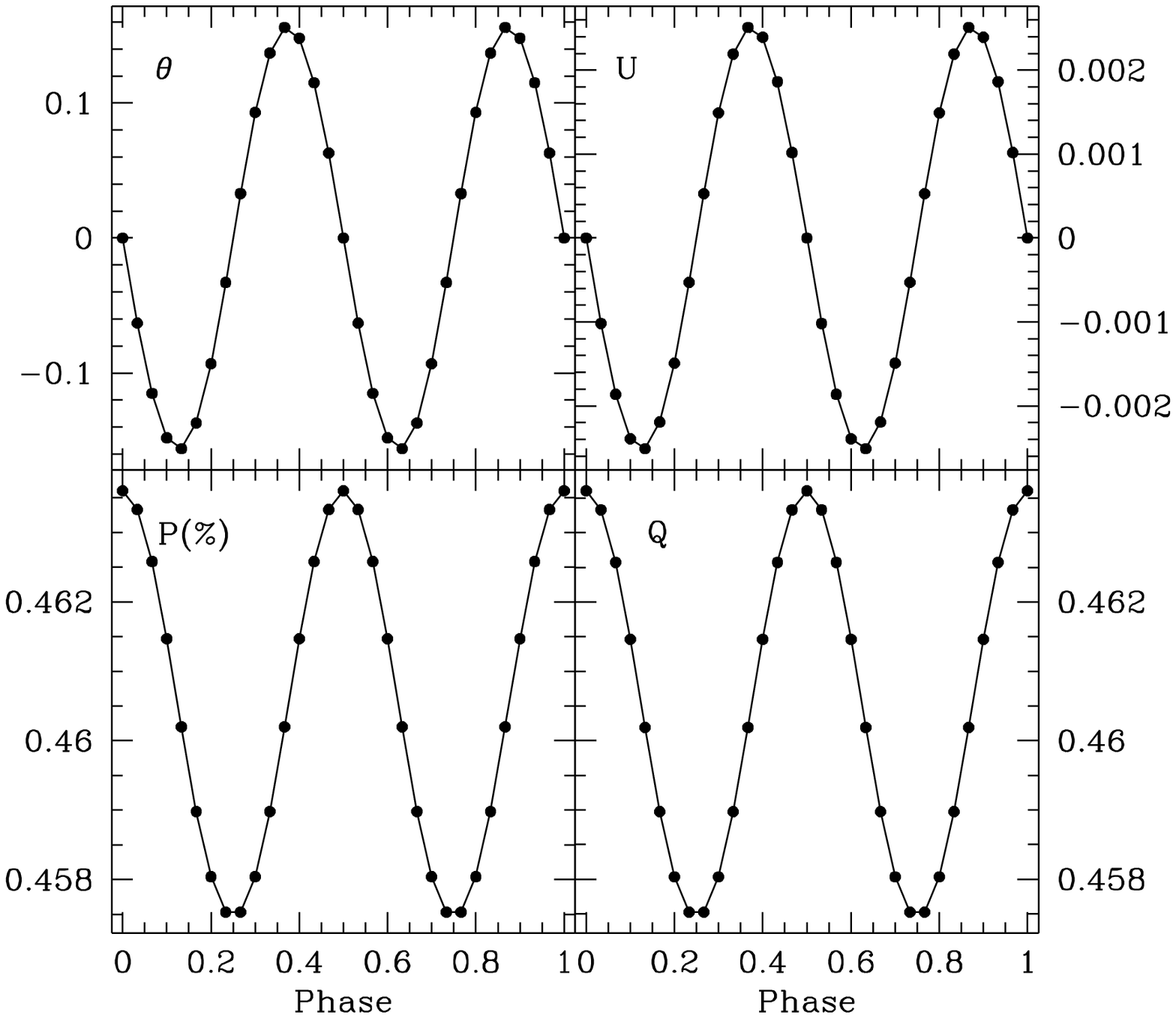}}
\figcaption[binary_AS_t01_a01.ps]{Typical simulation for a binary star
in a circumbinary envelope: polarization $P$, position angle $\theta$,
Stokes parameters $Q$ and $U$, as a function of the orbital phase
$\phi$. We used astronomical silicate grains of $0.1 \, \micron$, an
optical depth of 0.1, and an inclination of 60$^\circ$. The polarimetric
variations are purely double-periodic. \label{Fig-binary_AS_t01_a01}}

\newpage
\scalebox{0.75}{\includegraphics{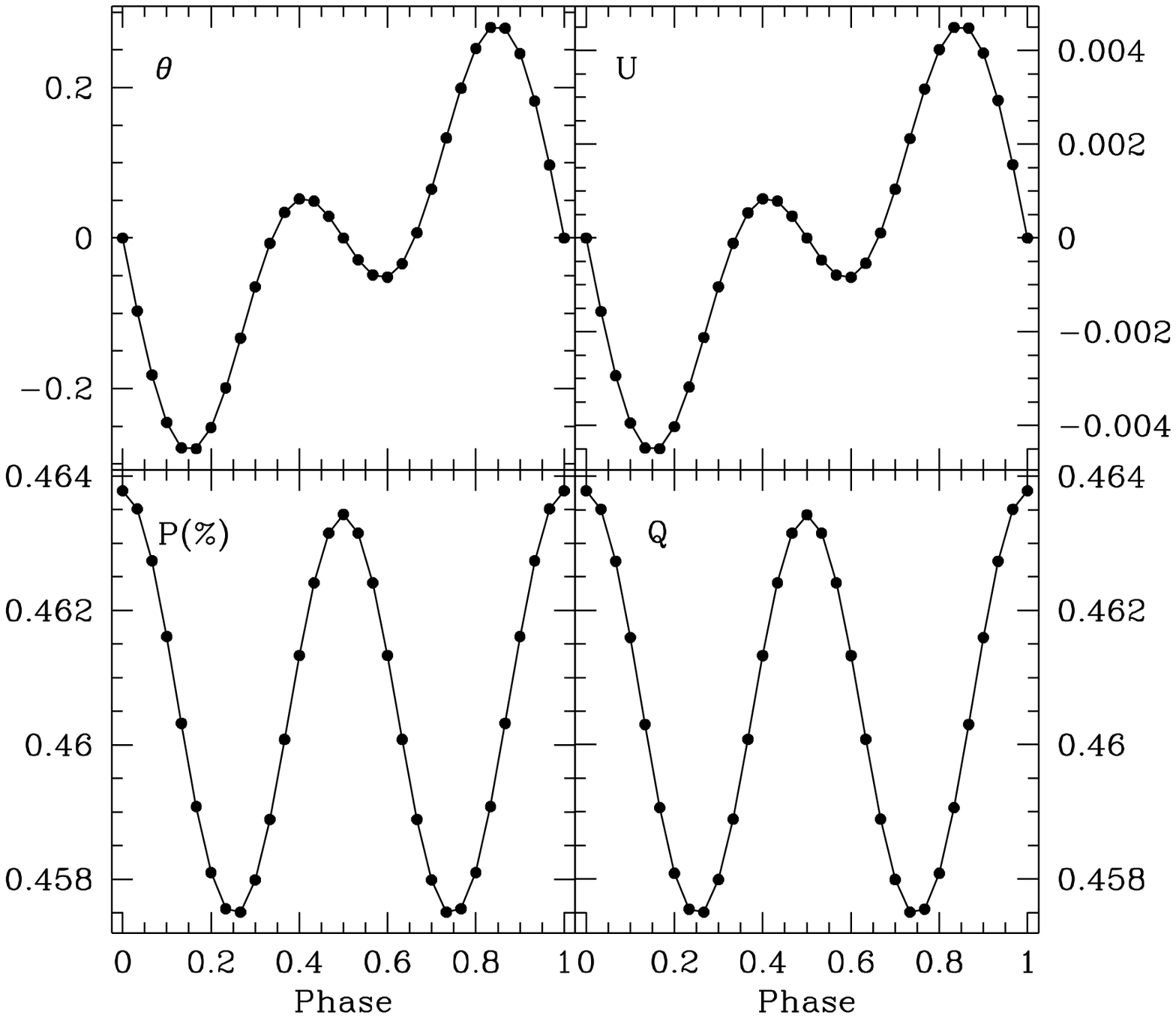}}
\figcaption[primary_AS_t01_a01.ps]{Same simulation as the previous
figure, taking only the primary star; the secondary is supposed to be
much fainter than the primary. In addition to the usual double-periodic
variations, single-periodic ones are clearly
seen. \label{Fig-primary_AS_t01_a01}} 

\newpage
\scalebox{0.75}{\includegraphics{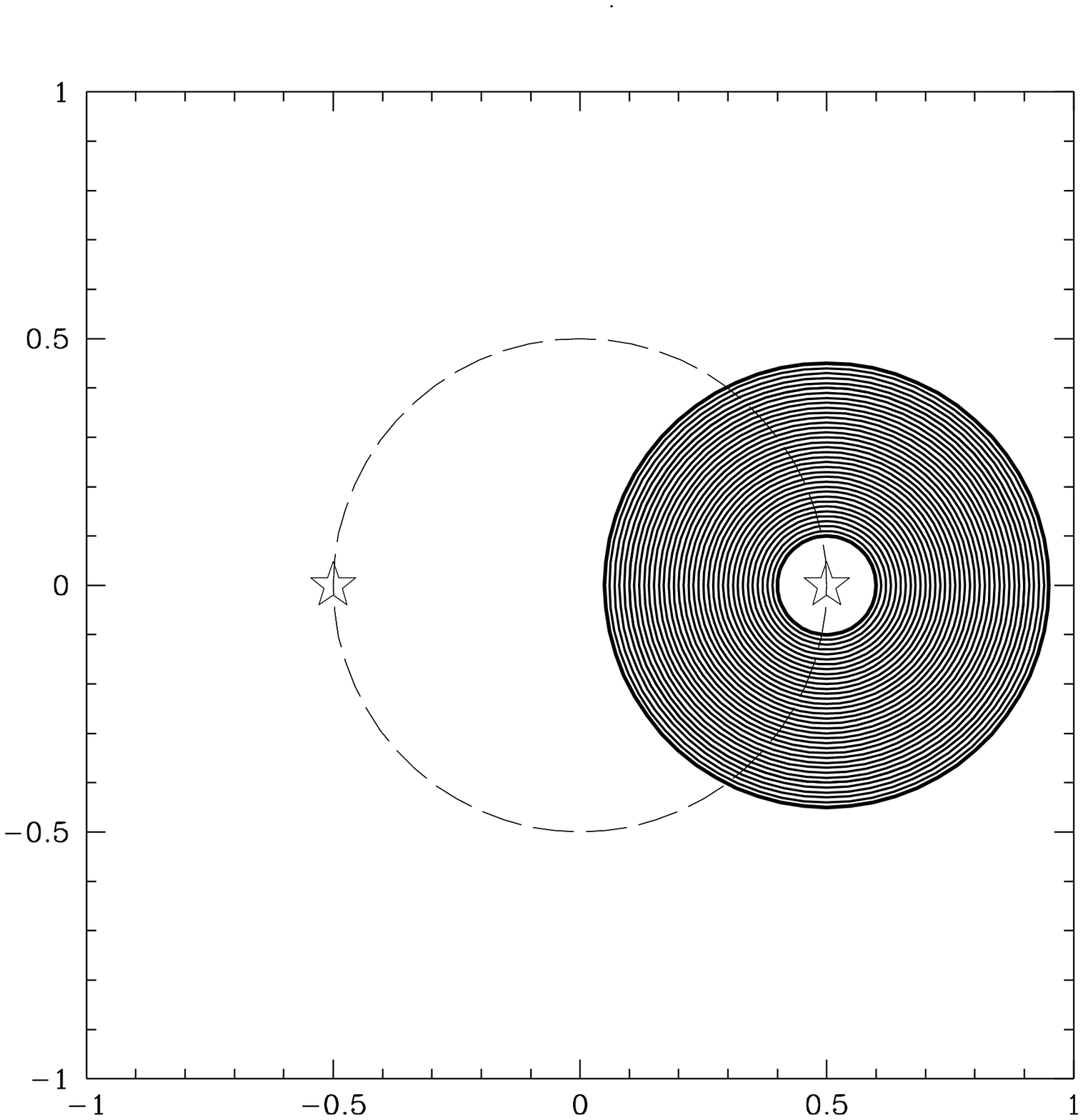}}
\figcaption[CS_45_1_5.ps]{Pole-on view of a binary system with only one
circumstellar envelope. \label{Fig-CS_45_1_5}}

\pagebreak

\begin{table*}
\begin{center}
\begin{tabular}{llll}
\tableline \tableline
Grain Composition & $\Re(m)$ & $\Im(m)$\tablenotemark{1} & $m$\\
\tableline
BE (amorphous carbon)\tablenotemark{2} & 2.240 & 0.781 & 2.37\\
Graphite \tablenotemark{3}	      & 2.524 & 1.532 & 2.95\\
Dirty ice \tablenotemark{4}	      & 1.33  & 0.09  & 1.34\\
Astronomical silicate \tablenotemark{5}& 1.715 & 0.030 & 1.72\\ 
\tableline
\end{tabular}                                                                 
\end{center}
\caption[Characteristics of dust grains used in the numerical
simulations]{Characteristics of dust grains used in the numerical
simulations. \label{Tab-grains}}
\tablenotetext{1}{The imaginary part of the complex refractive index $m$
represents the absorption of radiation by the grains.}
\tablerefs{(2) Rouleau \& Martin 1991; (3) Wickramasinghe \& Guillaume
1965; (4) Wickramasinghe 1967; Greenberg 1968; (5) Draine 1985}
\end{table*}

\pagebreak \clearpage

\begin{table*}
\begin{center}
\begin{tabular}{lllllll}
\tableline \tableline
$i_{\rm true}$ & & Electrons & Astron.  & Graphite &
      Amorphous  & Dirty Ice\\
           & &           & Silicates & & Carbon & \\
\tableline
  & $P_{\rm ave}$ (\%)     &0.794 &0.619 &0.299 &0.270 &0.189\\
90$^\circ$& $\Delta P$ (\%)       &0.0063&0.0048&0.0023&0.0021&0.0015\\
  & $P_{\rm ave}/\Delta P$&0.0079&0.0075&0.0077&0.0078&0.0079\\
\tableline
  & $P_{\rm ave}$ (\%)         &0.593 &0.461 &0.222 &0.201 &0.140\\
60$^\circ$& $\Delta P$ (\%)       &0.0079&0.0061&0.0029&0.0027&0.0018\\
  & $P_{\rm ave}/\Delta P$&0.0133&0.0132&0.0131&0.0134&0.0129\\
\tableline
  & $P_{\rm ave}$ (\%)        &0.395 &0.306 &0.147 &0.133 &0.093\\
45$^\circ$& $\Delta P$ (\%)       &0.0095&0.0073&0.0035&0.0032&0.0022\\
  & $P_{\rm ave}/\Delta P$&0.0241&0.0239&0.0238&0.0241&0.0237\\
\tableline
\end{tabular}                                                                 
\end{center}
\caption{Average polarization, $P_{\rm ave}$, and amplitude of the
polarimetric variations, $\Delta P$, produced by electrons and different
gain compositions, for different orbital inclinations.
\label{Tab-grains-e}} \end{table*}

\end{document}